# FANCA: In-Silico deleterious mutation analysis for early prediction of leukemia

Madiha Hameed Awan[1], Dr.Abdul Majiid,[*] Dr Asif Ullah Khan[2]

Pakistan Institute of Engineering and applied Sciences, Nilore Islmabad[1,2]

University of Gujrat[1]

**Abstract**

FANCA as novel biomarker is an antigen to leukemia, and it belongs to Fanconi Anemia Complementation Group (FANC) family. Non-synonymous Single Nucleotide Polymorphisms (nsSNPs) are an essential cluster of SNPs that caused alterations in encoded polypeptides. Mutations in the amino acid sequences of gene products caused leukemia.

Therefore, we conducted in-silico study to predict the FANCA related nsSNPs disease to analyze the association of nsSNPs with leukemia. The coding region of the dataset contains 100 nsSNPs and 24 missenses. In current study, the proteomic data is retrieved from the UniProtKB. Individual SNPs in the coding region of FANCA followed by the computation of deleterious mutation scores using four bioinformatics tools: PROVEAN, PolyPhen2, MuPro, and PANTHER. The 3-D structural and functional analysis is carried out and further found the multiple ligand sites that are useful for biological annotation.

We investigated SNPs into four amino acid groups of basic, polar, non-polar, and acidic. The results showed that **L1319Q** and **W209T** are the most deleterious mutations in the FANCA protein. The mutation L1319Q is changed into AA acidic group. The improper protein folding at location 1319 due to substitution of leucine into glutamine that loss of hydrophobic interaction. However, mutation W209T is changed into AA polar group at location 209 due to the substitution of tryptophan into threonine that are major causes of leukemia. It is inferred that these nsSNPs have a strong impact in the up-regulation of FANCA, which further elicit leukemia advancement.

Proteomic analysis for disease classification would be helpful in developing precision medicine. Our results would be beneficial for both researchers and practitioners in handling cancer-associated diseases related to FANCA.

# 1 Introduction

Leukemia is associated with many other cancers that initially start from the bone marrow and rapidly grow a large amount of anomalous blood cells. Leukemia comes from an immature blast of cells that resulted in the abnormality of leukocyte cells [1]. Based on cell origin, function and appearance, leukemia is divided into different types. Four major types are 1) acute lymphocytic leukemia (ALL), 2) acute myelocytic leukemia (AML), 3) chronic myelocytic leukemia (CML), and 4) chronic lymphocytic leukemia(CLL). The chronic type of leukemia commonly found in children and growth gradually. Whereas, the growth rate of acute leukemia is rapid compared to chronic type and gets worse quickly[1]. Timely detection of novel biomarkers and curative targets is an efficient way for the treatment of leukemia. The prediction of novel leukemia related to amino acids helps in identifying the protein sequence that develop cancerous cells[2].

FANCA Fanconi anemia, complementation group A, is a novel biomarker of Fanconi family of the protein sequence. It is based on 12 complementation groups belongs to FANC Family, comprising of eight homologous members FANCA, FANCB, FANCC, FANCE, FANCG, FANCH, FANCL, FANCM with 79-kilo bases (kb)[3], situated on chromosome 16(16q24.3). These eight family members do not share sequence similarities[5]. The composition of FANCA protein is based on1455 amino acids, and assembled together by a joint nuclear protein complex. It leads to multiple genetic irregularities of FANCA that become the cause of leukemia and other cancers. See **Figure 1**

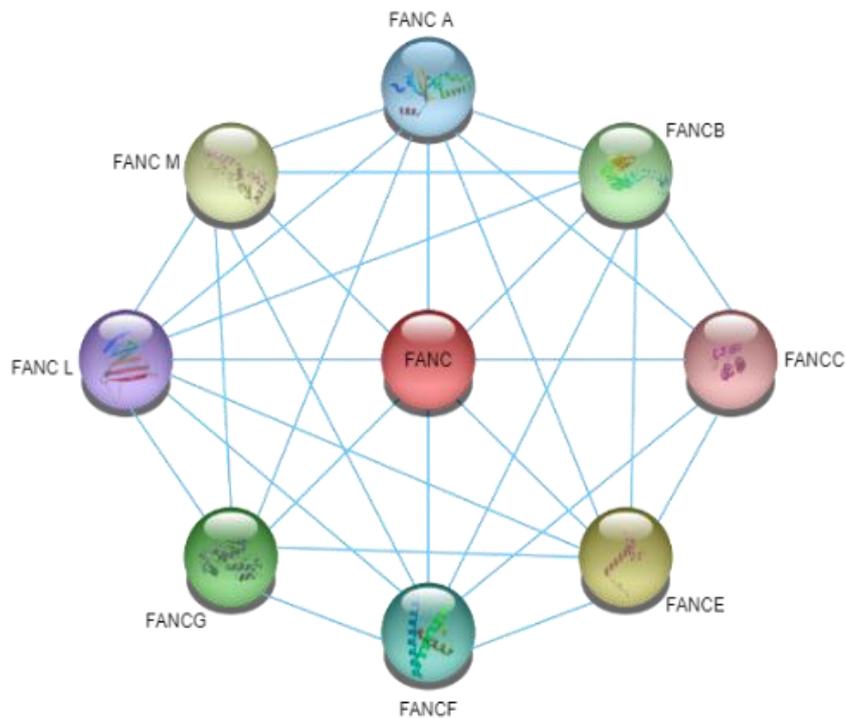

*Figure 1: Structure of FANC with its eight family members*

It is widely accepted that the variation of FANCA genes can be associated to diseases [6]. The proteins found in the human's body are: a) FAA, b) FACA, and c) FANCA. It is hypothesized that to activate as a post-replication restoration of a cell cycle checkpoint [7], FANCA protein repairs cross-link of DNA (Deoxyribonucleic acid ) and maintain a normal chromosome constancy that normalizes the variation of hematopoietic embryonic cells to developed blood cells**.** Non-synonymous single nucleotide polymorphism (nsSNP) is considered one of the most common types of interpretations[6]. Due to single-point mutation, Non-synonymous-SNP modification of an amino acid sequence in the protein has altered. SNPs are the mutual mutations inducing genomic alterations within the human's body. Previous studies showed that ~92% - 93% of human genes signify at least one SNP[7][8]. The mutations may be because of SNPs, duplications or deletions that affect multiple gene functions. The coding of non-synonymous (nsSNPs) are deleterious due to change in the physical and chemical properties of amino acids coding linked with the particular mutation[9][10]. The change in the nature of amino acids affects the protein translation polarity, stability, and accessibility. Disorder in the proteins eventually becomes the cause of its malfunctioning and molecular dynamics[11]. In recent era, the in-silico annotations are being used to evaluate the effect of SNPs on genomics and proteomics which subsequently helps us to early prediction of cancer and its grading. [12].

To address the challenges mentioned above, our contribution in this study are following

(i) Identify the deleterious nsSNP

(ii) Analyze the mutation on protein constancy to conclude whether the SNPs are deleterious or non-deleterious.

(iii) The deleterious mutation analysis is categorized into four different type of amino acid groups that help to understand their functional and structural property based biological function.

(iv) The multiple ligand sites are found that are useful for biological annotation.

This protein act as diagnostic biomarkers in several inherited human diseases such as neural abnormalities, vascular conditions[16], tumors [17], etc. In current work, the proteomic data is obtained from the UniProtKB. We found the coding region contains 100 non-synonymous single nucleotide polymorphisms (nsSNPs) and 24 missense SNPs as deleterious and divide them into four amino acid groups. It is inferred that nsSNPs could help the up-regulation of FANCA[18]. In this study, the main contribution is the exploration of a new dimension for early prediction of Leukemia using deleterious mutation information in amino acid sequences.

## 2 Proposed Framework

The proposed framework of deleterious mutation identification and structural prediction is reported in **Figure 2**. The input data of FANCA protein, related to amino acid composition variations, is obtained from UniprotKB. The Basic Local Alignment Search Tool (BLAST) algorithm is employed to deduce functional and evolutionary relationships between the sequences of FANCA. Homologs of sequence queries are identified using the BLAST[19]. FANCA has ten transcripts in the database, but we chose only known transcripts. Non-synonymous SNPs of the recognized transcripts are selected. Afterwards, various tools are used to ascertain the functional and structural effects of all non-synonymous SNPs. These tools are based on diverse procedures to find the impact on protein sequences.

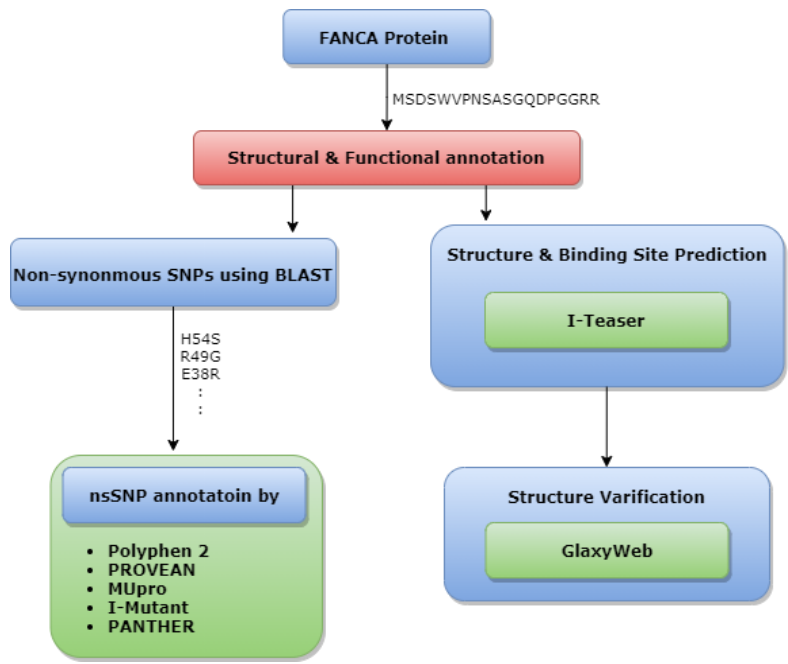

*Figure 2: proposed Framework for deleterious mutation identification and structural prediction*

In the proposed framework, for SNPs annotations, five different tools PolyPhen2[20], PROVEAN[21], Mupro[22], I-Mutant[23] and PANTHER [24]are employed for different purposes. For example, PolyPhen2 predicts the damage of missense mutations. PolyPhen2 predicts the damage of missense mutations. It uses iterative greedy algorithm to identify the sensitivity and specificity score of that mutation which helps to find out the severity of damage.

On the other hand, Protein Variation Effect Analyzer (PROVEAN) tool uses the location-specific score approach that takes Protein sequence and amino acid variations as input. A cutoff value -2.5 is set for the given binary prediction to achieve balanced accuracy. Amino acid substitutions with a value less than the threshold is considered deleterious. PROVEAN tool uses CD-HIT clustering algorithm for BLAST run. It returns the identity of 75% global sequence. The top 30 identified sequences are closely related to the groups of sequences from the supporting domain that are used to produce the prediction results. A delta orientation score is calculated for each supportive sequence[26]. PROVEAN calculates the prediction score by employing the following equation.

$$\text{Score} = \Delta \text{ (Q or S)}, \qquad (1)$$

where, Q indicates the query sequence used for the score calculation. However, symbols S represent the collected protein sequences. The symbol V indicates the actual variant of specific protein sequence. The final PROVEAN score is generated by averaging the accumulated score within and across the clusters. If the PROVEAN score is ≤ -2.5, the protein variant is predicted as "deleterious". Otherwise, the variant is expected to be "neutral".

Mupro tool is used to ascertain the protein stability prediction from a sequence of a single site mutation[27]. This tool is empowered with Support Vector Machine (SVM) and Neural Network algorithms to predict the increase/decrease stability with a confidence score of single-site mutation of amino acid. The original amino acid is given with mutation location and mutated amino acid of a protein sequence to implement the algorithm.

To calculate the consistency of nsSNPs based on function and structure of target protein, I-MUTANT is another tool used for prediction that is based on SVM algorithm[28][29]. I-Mutant sets Delta-Delta Gibbs (DDG) free energy value within the range of -0.5 to 0.5. This value is an indicator to identify how much a single site mutation is affecting protein consistency. The more negative value result in higher decrease of stability[30].

On the other hand, PANTHER tool finds mutation based on algorithms such as statistical modelling, Multiple Sequence Alignment (MSA), and Hidden Markov Model (HMM)[31]. It is a suite of tools to identify query sequence functions and analyzes large-scale experimental data with several statistical tests. Biologists widely use PANTHER to identify protein mutation stability.

In the proposed framework, Iterative Threading Assembly Refinement (I-Tasser) is used for 3D structure prediction of FANCA and its biological functions based on amino acid sequences[32][33]. Functionality of I-Tasser is based on three steps: a) Iterative structure assembly, b) structural template identification, and c) structure-based function annotation. The confidence score (C-Score) is the selection criteria of the model[34]. The high value of C-score indicates the better model. FANCA structure evaluated for both wild-type and mutated protein [35]. The protein model is refined using GalaxyWeb tool. It takes Protein Data Bank (PDB) format as input file for refinement of models[36]. GalaxyWeb employed Z score value of HHsearch results. HHSearch algorithm is associated with three well-known search BLAST, PSI-BLAST, HMMER[36][37]. These model are further used to compare different databases to identify the pairwise sequences, and to further improve the quality of global and local structure quality[38]. The re-ranking is the accumulated result of Z-score. It can be obtained from HHsearch sequence score Zss and Zseq.

$$S_{hh} = Z_{seq} + w\, Z_{s\text{-score}} \qquad (2)$$

MetaServer approach named as COACH, based on the combination of ligand-binding site and multiple function annotation, is the base of COFACTOR algorithm, TM-SITE & S-SITE program used to find out ligand binding sites. The particular ligands with higher C-score highlighted more confidence to specify consistent prediction.

## 3. Results and Discussion

In this section, first, we will analyze the mutation results derived using above mentioned five S/w tools. Identify the deleterious nsSNP. The multiple ligand sites are found that are useful for biological annotation. The deleterious mutation analysis is categorized into four different type of amino acid groups that help to understand their functional and structural property based biological function. The multiple ligand sites are found that are useful for biological annotation

### 3.1 SNPs analysis

The damage of missense mutations computes the likelihood of submitted variants based on obtaining results. A higher value with high sensitivity and specificity indicates the higher damaging effects of missense mutation. In Figure 3, the black vertical line denotes the Q387V mutation predicted with possible damage score, sensitivity, and specificity values of 0.9468, 0.80, and 0.95, respectively. Figure shows the range of score in between 0-1. Table 2 indicates the higher deleterious prediction score of -4.513 for variant Q387V.

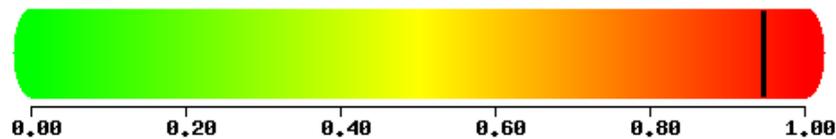

*Figure 3: Damaging prediction score for Q387V mutation*

The location-specific score, for the binary prediction, is computed using protein sequence and amino acid variations are taken as input with the cutoff value of -2.5. The balanced accuracy is achieved with 27 number of clusters and 64 supporting sequences. The mutations L1319Q and W209T computed as deleterious with score -2.707 and -6.856, respectively.

Delta-Delta Gibbs(DDG) free energy value within the range of -0.5 to 0.5 is used to identify how much a single site mutation is affecting protein consistency. The more negative value of -1.87 result in less stability, with pH level 25, indicate the alarming condition of protein mutation.

### 3.2 3D structure prediction

For 3D structure prediction of FANCA is predicted and biological functions based on amino acid sequences is also analyzed. FANCA structure evaluated for both wild-type and mutated protein. FANCA best model for mutated protein have C-score = 0.96, Estimated TM-score = 0.84±0.08, Estimated RMSD = 7.5±4.3Å. The best model for wild-type protein with C-score = -1.23, Estimated TM-score = 0.56±0.15, Estimated RMSD = 9.7±4.6Å. The Figure 4 (a) and (b) indicates mutated and wild-type of FANCA protein Structure, respectively.

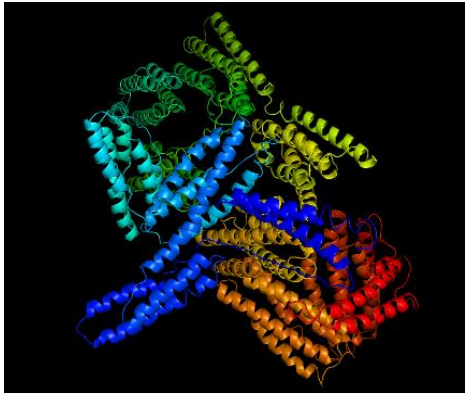
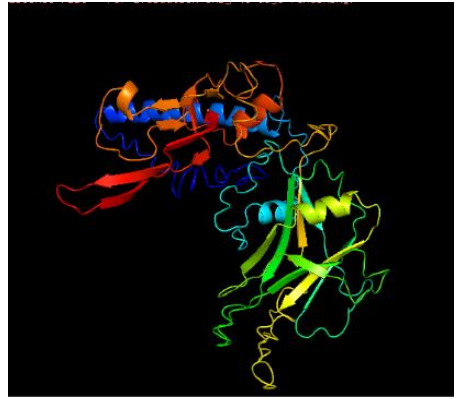

(a)                  (b)

*Figure 6: FANCA Protein Structure; (a) Mutated protein structure and (b) Wild-type protein structure*

### 3.3 Protein Model Refinement

The protein model is refined for its SNPs annotation of FANCA protein. Z score value of HHsearch results are taken. HHSearch algorithm is associated with three well-known search BLAST, PSI-BLAST, HMMER. Figure 7 (a) - (e) indicates the refined models respectively.

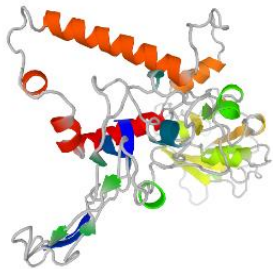
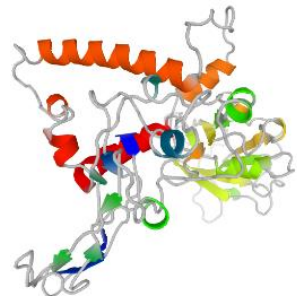

(a)                  (b)

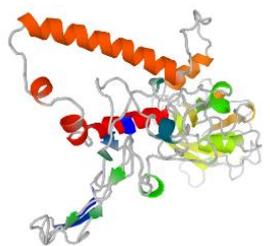
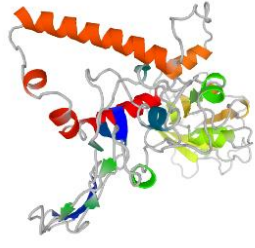

(c)                  (d)

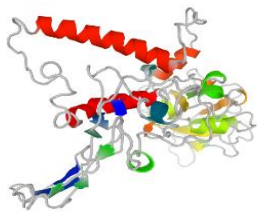

(e)

*Figure 7: FANCA protein refined structure model(a) to model(e)*

Table 1 demonstrates the numerical values of all refined models obtained from global distance test. Root means square and all relevant results are used to refine the FANCA structure. RMSD of a ligand is necessary to check if it is stable in the active sites and identify possible binding modes. All the value associated with the refined model of FANCA protein structure initial values are given in the first row and the comparison of obtain value shows that model c in Figure 7 has the highest RMSD of value 0.501that indicates that model c have strong binding site. This inferred that model c is good for precision medicine implementation and help in leukemia treatment. Same as GDT-HA, Mol probility, clash score, poor rotamers and rama favored have their influence in the protein structure refined model. Variation in their values effects the models structure accordingly that is mention in table below in detail.

*Table 1: Refined Models of FANCA protein structure*

| Models | GDT-HA | RMSD | Mol Probity | Clash score | Poor rotamers | Rama favored |
|---|---|---|---|---|---|---|
| Initial | 1.0000 | 0.000 | 3.498 | 146.7 | 0.0 | 66.3 |
| a | 0.9096 | 0.499 | 2.059 | 13.7 | 0.2 | 93.7 |
| b | 0.9122 | 0.000 | 2.078 | 13.9 | 0.3 | 93.5 |
| c | 0.9096 | 0.501 | 2.091 | 14.1 | 0.3 | 93.3 |
| d | 0.9137 | 0.000 | 2.068 | 14.5 | 0.6 | 94.0 |
| e | 0.9098 | 0.500 | 2.074 | 13.7 | 0.2 | 93.4 |

### 3.4 FANCA protein SNP annotation

Further studies, nsSNPs among all other variants were chosen. More than 100 mutations with four amino acid groups detected in FANCA protein, 24 of them are identified found deleterious.

Group 1 Basic Amino acid group mutations are E38R, L406R, P597R. Group 2 Acidic amino acid mutations are G66Q, V77D, A403D, L587E, L1319Q and V566N.Group 3 Polar amino acid group mutations are H54S, R49G and W209T and the last Group 4 Non polar amino acid group mutations are Y35M, C625A, G638I, L43A, V378I, V397L, P401A, L406R, F1108A, H780A, I573V, E63L, Q387V. Theses nsSNPs cause variation within target protein sequences that progressively affect the protein functions. Table 2 reports the details of four amino acid group that were analyzed as deleterious.

*Table 2: Identified deleterious mutation with score using software tools.*

| S# | AA Group | Deleterious Mutation | PolyPhen2 prediction | PolyPhen2 score | PROVEAN prediction | PROVEAN cutoff | MUpro | I-Mutant | PANTHER |
|---|---|---|---|---|---|---|---|---|---|
| 1 | Basic | E38R | POSSIBLY DAMAGING | 0.946 | Neutral | -2.055 | DECREASE | DECREASE | Probably Benign |
| 2 | Basic | L406R | POSSIBLY DAMAGING | 0.891 | Deleterious | -3.037 | DECREASE | DECREASE | Probably Damaging |
| 3 | Basic | P597R | PROBABLY DAMAGING | 0.995 | Deleterious | -5.015 | DECREASE | DECREASE | Probably Damaging |
| 4 | Acidic | G66Q | PROBABLY DAMAGING | 0.961 | Neutral | -1.85 | DECREASE | DECREASE | Probably Benign |
| 5 | Acidic | V77D | BENIGN | 0.039 | Neutral | -0.395 | DECREASE | DECREASE | Probably Benign |
| 6 | | A403D | PROBABLY DAMAGING | 0.985 | Deleterious | -3.434 | DECREASE | DECREASE | Probably Benign |
| 7 | | L587E | PROBABLY DAMAGING | 1.00 | Deleterious | -5.288 | DECREASE | DECREASE | Probably Damaging |
| 8 | | L1319Q | PROBABLY DAMAGING | 1.00 | Deleterious | -2.707 | DECREASE | DECREASE | Possibly Damaging |
| 9 | | V566N | POSSIBLY DAMAGING | 0.789 | Deleterious | -3.113 | DECREASE | DECREASE | Probably Damaging |
| 10 | | H54S | BENIGN | 0.42 | Deleterious | -0.324 | DECREASE | DECREASE | Possibly Damaging |
| 11 | Polar | R49G | BENIGN | 0.208 | Neutral | -2.281 | DECREASE | DECREASE | Probably Benign |
| 12 | | W209T | PROBABLY DAMAGING | 1.00 | Deleterious | -6.856 | DECREASE | DECREASE | Probably Damaging |
| 13 | | Y35M | POSSIBLY DAMAGING | 0.794 | Neutral | -1.441 | DECREASE | DECREASE | Probably Benign |
| 14 | | C625A | PROBABLY DAMAGING | 0.963 | Deleterious | -6.841 | DECREASE | DECREASE | Probably Damaging |
| 15 | | G638I | PROBABLY DAMAGING | 0.992 | Deleterious | -3.256 | DECREASE | DECREASE | Probably Damaging |
| 16 | | L43A | POSSIBLY DAMAGING | 0.929 | Neutral | -2.254 | DECREASE | DECREASE | Possibly Damaging |
| 17 | | V378I | POSSIBLY DAMAGING | 0.908 | Neutral | -0.647 | DECREASE | DECREASE | Probably Damaging |
| 18 | Non-polar | V397L | POSSIBLY DAMAGING | 0.78 | Neutral | -0.578 | DECREASE | DECREASE | Probably Damaging |
| 19 | | P401A | BENIGN | 0.18 | Deleterious | -2.931 | DECREASE | DECREASE | Possibly Damaging |
| 20 | | Q387V | POSSIBLY DAMAGING | 0.946 | Deleterious | -4.513 | INCREASE | DECREASE | Probably Damaging |
| 21 | | F1108A | PROBABLY DAMAGING | 1.00 | Deleterious | -4.36 | DECREASE | DECREASE | Possibly Damaging |
| 22 | | H780A | PROBABLY DAMAGING | 0.999 | Deleterious | -6.244 | DECREASE | DECREASE | Probably Damaging |
| 23 | | I573V | BENIGN | 0.429 | Neutral | -0.514 | DECREASE | DECREASE | Probably Damaging |
| 24 | | E63L | PROBABLY DAMAGING | 0.999 | Deleterious | -3.932 | DECREASE | DECREASE | Possibly Damaging |

### 3.5 Chemical analysis of mutated AA

Chemical properties divide AA into four group Basic amino acid group, Polar amino acid group, Non-polar amino acid group and acidic amino acid group. The deleterious mutation is categorized into four different type of amino acid groups that help to understand their functional and structural property based biological function.

### 3.5.1 Transition to Basic AA:

There are three variant lies in Basic group of amino acid properties. First mutation in **E38R** glutamic acid is substituted with arginine that changes its group from acidic to basic amino acid property; this results in variation in charge of ligands and/or other residues. This mutation become the cause to abnormal folding[44]. In transformation, **L406R** leucine substituted with arginine at location 406. This deleterious mutation increases its iconic bond that becomes the cause of malfunctioning and disturbs its particular conformation [58]. Third mutation **P597R** here proline replace with arginine at location 597 with and overall charge of +1 at physiological pH. This deleterious mutation causes abnormal folding the functionality of core protein with the damaging score of 0.993 [63].

### 3.5.2 Transition to Acidic AA:

Portion of side chain in amino acid group that contain negative charge at certain pH value is acidic group and there are 6 out of 24 mutation lies in acidic group. **G66Q** Glycine amino acid is substituted with glutamine at location 66.This mutation loss the polarity into acidic property that causes the loss of charge; hence, the molecular interaction is also lost[49]. In mutation, however, in mutation **V77D** Valine amino acid substituted with aspartic acid at location 77 that become the cause of external interaction. This mutation is also cause of loss hydrophobicity that reduces oxygen affinity[50]. Mutation **A403D** alanine substituted by aspartic acid at location 403. This mutation causes the loss of hydrophobic interaction. It also reduces the toxicity and activity of the enzyme[57]. In the third mutation **L587E** here leucine is substituted with glutamic acid at location 587.This mutation becomes the cause to loss of core protein hydrophobic interaction[62]. Mutation **L1319Q** leucine replaced with glutamine at location 1319. This mutation becomes the reason to abolish the function and proper binding of protein by entering acidic group [68]. And the last mutation **V566N** valine substituted with asparagine at location 566. This mutation becomes the cause of destabilization of local conformation and a significant decrease in some specific activities[59][60].

### 3.5.3 Transition to Polar AA:

Uncharged amino acid group where the side chain in this group is possess a spectrum of a functional group. First of its mutation where **H54S** histidine is replaced with serine at location 54 and change its properties from basic to acidic that causes low surface tension at the core of protein would be vanished[41]. However, in mutation, **R49G** arginine is substituted by glycine at location 49. Arginine lies in basic group of amino acid and after this mutation serene comes in group of acidic activate their side chain that have carboxylic acid group whose pKa's are low enough to lose proton and becoming negative charged in the process. The transformation would break hydrogen bonds and/ or result in improper folding[42][43]. While mutation **W209T** Amino acids tryptophan is substituted with threonine at location 209, which becomes the cause of loss of hydrophobic interactions[51].

### 3.5.4 Transition to Nonpolar AA:

Non polar side chains consist mainly of hydrocarbon. Any functional groups they contain are uncharged at physiological pH and are incapable of participating in hydrogen bonding that loss the polarity of the amino acid group and raise the hydrogen bonds in it. Like in mutation **Y35M** tyrosine is replaced with methionine that loss the charged at physiological pH of that causes the failure of hydrogen bonds and disturbs correct folding. Nonpolar property have negative charge comes increased [45]. **C625A** cysteine substitute with alanine at location 625.This damaging mutation effects on the ligands charge[64]. However, **G638I** Amino Acid glycine is replacing with leucine at location 638.This deleterious mutation become the cause to loss of positive charge[65].

Mutation **L43A** occur by changing leucine into alanine at location 43. This mutation can interrupt its particular conformation and revoke its function by stay still nonpolar group its carboxylic acid remains same.[46]. In mutation, **V378I** valine amino acid replaces with isoleucine at location 378, which becomes the cause of mutation and disrupts local conformation, leading to loss of interaction[52]. In mutation, **V397L** valine amino acid returns with leucine at location 397, which becomes the cause of upsetting the region's stability and causes the loss of interaction[54]. Mutation **P401A** proline substituted by alanine at location 401. The substitution can interrupt its particular chemical structure and eliminate its function. Proline mutation into alanine affects the reputed integral membrane protein segments 6 and 10 responsible for glucose transportation GLUTl [55][56].

In mutation, **F1108A** phenylalanine substituted with alanine at location 1108. This deleterious mutation become the causes of loss of hydrophobic interaction[67]. The third basic group mutation is also **H780A** histidine substation with alanine at location 780, this mutation changes in nonpolar group and becomes the reason to disturb domain by losing hydrophilicity[66]. Mutation **I573V** Isoleucine substituted with valine at location 573 becomes the causes the loss of interaction. Secondary structures of α helice is not preferred by this type of residue [61]. In mutation **E63L** Glutamic acid is being mutated in leucine at location 63. It changes its acidic properties into nonpolar that activate its carboxylic acid. It becomes the cause of metabolism functionality and may also cause intellectual disability[47][48]. Mutation **Q387V** glutamine replaced with valine at location 387 causes the loss of proper folding and conduce loss of hydrogen bond[53].

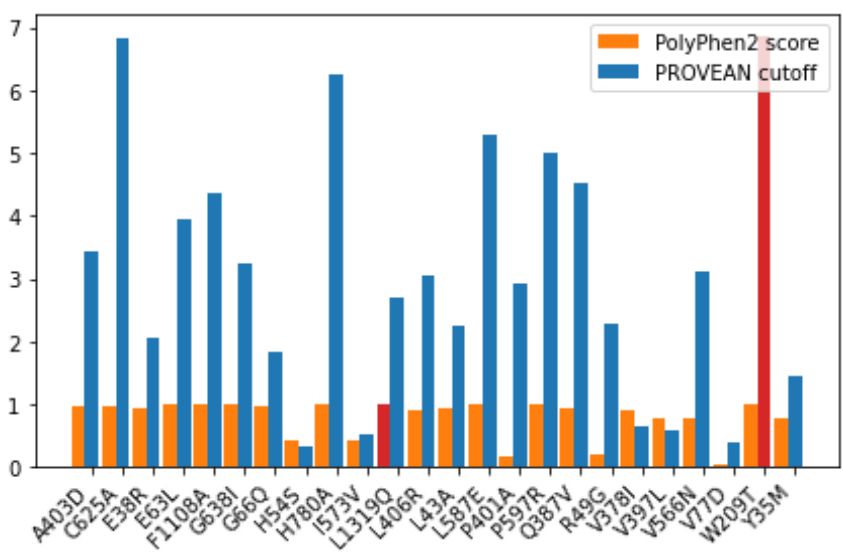

*Figure 6: Most Deleterious Mutation Scores using Polyphen, PROVEAN. * In order to better visualize, the negative values are plotted as a positive one.*

**Figure 6** Demonstrate the Polyphen and PROVEAN score of all mutations that occur in the target protein. Results are increasing gradually according to their score level. From Table 2, it is observed that L1319Q and W209T is most deleterious mutation in FANCA protein.

## 3.6 Ligand Binding Site prediction in FANCA protein

Ligand (L) binding sites have a significant role in protein functionality. Mutation in protein sequence disrupts the collaboration between protein ligands and transmembrane. Figure 8, shows FANCA protein structure highlighted (a) binding site H492 (b) binding site R517 (c) binding site E484 (d) binding site P565, (e) binding site S175 (f) binding site W183 (g) binding site Y448 (h) binding site k467.

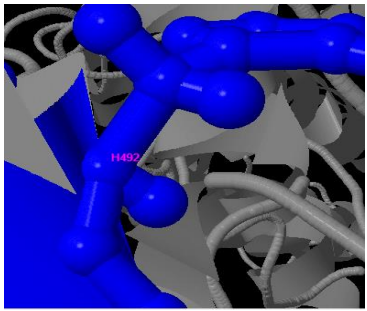

(a) Ligand binding site H492, C-score=0.03, Cluster-size=2, L-Name=R1P

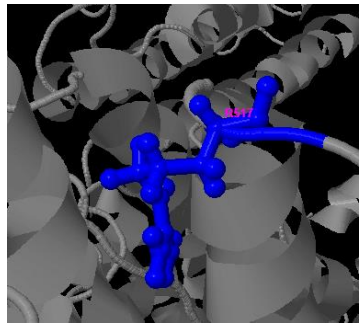

(b) Ligand Binding site R517, C-score=0.03, Cluster-size=2, L-Name=R1P

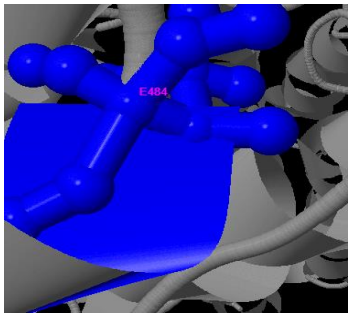

(c) Binding site E484, C-score=0.03, Cluster-size=2, L-Name= TRE

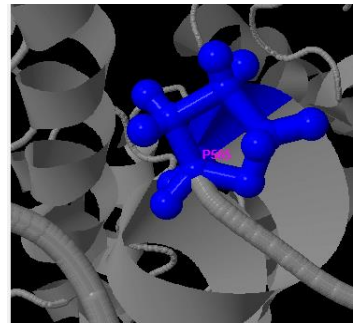

(d) Binding site P565, C-score=0.03, Cluster-size=2, L-Name= TRE

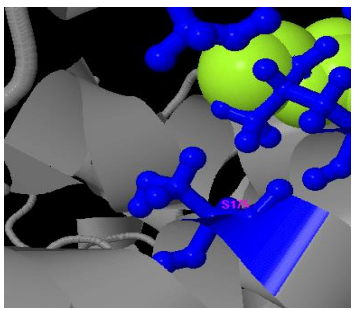

(e) Binding site S175, C-score=0.03, Cluster-size=2, L-Name= K85

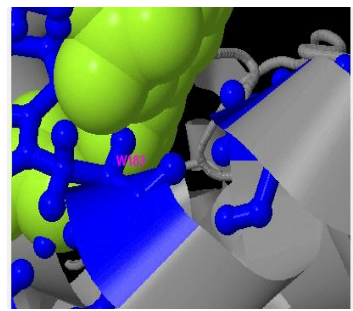

(f) Binding site W183, C-score=0.03, Cluster-size=2, L-Name= TRE

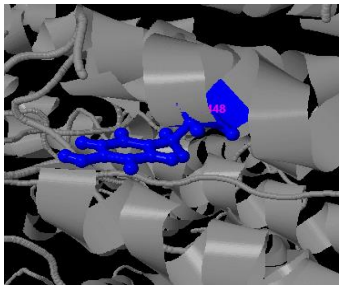 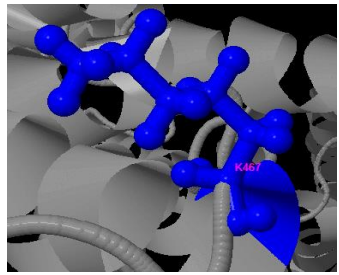

(g) Binding site Y448, C-score=0.02, Cluster-size=1, L-Name= IMD

(h) Binding site k467 ,C-score=0.02, Cluster-size=1, L-Name= IMD

*Figure 8 (a-h): FANCA Ligand Binding Sites with C-score, cluster-size, L-Name*

The multiple ligand sites are found that are useful for biological annotation. In other hand, Both COFACTOR and COACH algorithms used for the target protein structure prediction. However, COFACTOR deduces the functions of protein such as ligand-binding sites, enzyme commission (EC), and gene ontology (GO) using structure comparison and protein-protein networks. MetaServer approach named as COACH, based on the combination of ligand-binding site and multiple function annotation, is the base of COFACTOR algorithm, TM-SITE & S-SITE program.

The particular ligands with higher C-score highlighted more confidence to specify consistent prediction. The ligand R1P has a higher C-score than the remaining ligands, and its potential binding sites. The above mention table shows that ligand binding sites 492 and 517 have more C-score value of 3 than others. This mean it is stronger site for residue binding. These binding sites useful in precision medicine.

## 6. Conclusion

In this modern era, proteomic analysis for disease classification related to single nucleotide polymorphisms (SNPs) is tremendously inspiring. Bioinformatics helps to reduce the genotyping cost that increases omics association studies. We conducted in-silico study to predict the FANCA related nsSNPs disease and multiple ligand sites find out to analyze the biological annotation of ligand binding sites. We analyze the association of nsSNPs with leukemia. In this study, 24 mutations are found that have been predicted as deleterious by all structural and sequential prediction bioinformatics tools. The deleterious mutation analysis is carried out from four different type of amino acid groups Basic, Polar, Non-Polar, Acidic to understand their physiochemical property. It is inferred that L1319Q and W209T both changes their amino acid groups one moves into acidic group where other transform into polar group that are either charged at physiological pH are major causes of leukemia. The results showed that these nsSNPs affects the functional and structural mechanism of FANCA, which plays a significant role in acute myeloid leukemia. It increases the chance of mitotic cell division, specifically in the bone marrow. The ligand-binding site prediction have a vital role in the precision medicine for leukemia. So, this study will be a valuable addition to the research world to predict the consequence of nsSNPs of FANCA in the up-regulation of leukemia. It is anticipated that the prediction of deleterious mutations in the genomic functionality helps in the early detection of leukemia. The proposed study would also help to develop precision medicine in the field of drug discovery.